\documentclass[a4paper,11pt]{article}
\usepackage{graphicx}
\usepackage[latin1]{inputenc}
\usepackage{amsfonts}
\usepackage{amssymb}
\usepackage{amsmath}
\usepackage{a4wide}
\usepackage{hyperref}
\usepackage{cite}
\usepackage{caption}
\usepackage{mdwlist}
\usepackage{color}
\usepackage{ulem} 

\begin{document}

\begin{flushright}
TIFR/TH/16-13
\end{flushright}
\vskip24pt

\begin{center}
{\Large Exploring the Inert Doublet Model through the dijet plus missing transverse energy channel at the LHC}\\[10mm]

\noindent
{\large P. Poulose\footnote{poulose@iitg.ernet.in},  Shibananda Sahoo}\footnote{shibananda@iitg.ernet.in}\\Department of Physics, Indian Institute 
of Technology Guwahati, \\Assam 781039, India.\\ [5mm] 
{\large K. Sridhar\footnote{sridhar@theory.tifr.res.in} }\\ Department 
of Theoretical Physics, Tata Institute of Fundamental Research, \\ Homi Bhabha
Road, Mumbai 400 005, India
\end{center}

\begin{abstract}
In this study of the Inert Doublet Model (IDM), we propose that the dijet + 
missing transverse energy channel at the Large Hadron Collider (LHC) will be an 
effective way of searching for the scalar particles of the IDM. This channel 
receives contributions from gauge boson fusion, and $t-$channel production, 
along with contributions from $H^+$ associated production. We perform the 
analysis including study of the Standard Model (SM) background with assumed systematic uncertainty, 
and optimise  the selection criteria employing suitable cuts on the kinematic variables to 
maximise the signal significance.  We find that with high luminosity option of the LHC, this 
channel has the potential to probe the IDM in the mass range of up to about 
400 GeV, which is not accessible through other leptonic channels. In a scenario with light dark matter of 
mass about 65 GeV, charged Higgs in the mass range of around 200 
GeV provides the best possibility with a signal significance of about $2\sigma$ at an integrated luminosity of about 3000 fb$^{-1}$.
\end{abstract}

\newpage

\section{Introduction}
\hskip .8cm 
The discovery of Higgs boson by the ATLAS and CMS collaborations of the LHC
 \cite{Aad:2012tfa,Chatrchyan:2012ufa} has definitely put the spotlight of particle physics research on Higgs phenomenology. While all measurements so far indicate that the new particle is indeed a Higgs boson, compatible with that predicted by the Standard Model (SM) of particle physics, detailed questions about the
exact nature of the Higgs potential and the coupling of the Higgs particle
to other SM particles need to be investigated. This information, along with the popular reasoning that the SM Higgs mechanism is only an effective description to understand Electroweak Symmetry Breaking (EWSB), of a more fundamental 
high-energy theory has led researchers to study the implication of many variant models. It is believed that the Standard Higgs mechanism with only one physical scalar is too minimalistic, and in reality, there could be more than one Higgs field sharing the responsibility of EWSB. Such multi-Higgs models are also motivated by other drawbacks of the SM. For example, low-energy supersymmetric models (Minimal Supersymmetric Standard Model - MSSM) proposed as a remedy to the hierarchy problem requires two doublet Higgs fields, resulting in the physical spectrum with five more scalars, three of which are neutral. The two-Higgs Doublet Models (2HDM) without supersymmetry has also been a popular theoretical option beyond the minimal Higgs mechanism  proposed by the SM. The issue of dark matter, required by astrophysical observations but for which we lack a suitable candidate in the SM, is another important reason to attempt to go beyond the SM and, in these attempts, it is often the minimalism of the scalar sector of the SM that is sacrificed. 

The 2HDM with one of the doublet fields not having any direct (at the level of the Lagrangian) interaction with the SM particles, except the gauge particles, is a promising candidate model in this regard. This is achieved by the imposition of a $Z_2$ symmetry under which one of the doublets is odd, while all other fields are even. Such an Inert Doublet Model (IDM) \cite{Deshpande:1977rw} would have the Higgs phenomenology, quite different from that of the SM as well as the MSSM or the usual 2HDM scenarios. For example, in the physical Higgs boson sector, all neutral scalars except one are odd under the $Z_2$ symmetry and are, therefore, always produced in pairs. This also means that the lightest of these cannot decay, and thus could be a candidate for dark matter. Adding a $Z_2$-odd right-handed neutrino to this model can also generate small neutrino masses radiatively\cite{Ma:2006km}, and to generate leptogenesis \cite{Ma:2006fn}, ideas which are followed up in further studies \cite{Krauss:2002px,Kubo:2006yx, Sierra:2008wj,Suematsu:2009ww,Suematsu:2010gv,Suematsu:2011va, Kashiwase:2012xd,
  Kashiwase:2013uy, Chakrabarty:2015yia}. The model is shown to be helpful in explaining the LEP-paradox \cite{Barbieri:2000gf, Barbieri:2006dq,Casas:2006bd}, and could also generate EWSB at one-loop level through Coleman-Weinberg mechanism \cite{Hambye:2007vf}. 

\vskip .2cm
\hskip .8cm With its interesting Higgs phenomenology, the IDM has been studied in the context of the Large Hadron Collider (LHC) \cite{Cao:2007rm, Dolle:2009ft,Miao:2010rg,Gustafsson:2012aj,Arhrib:2012ia,Krawczyk:2013wya,
Krawczyk:2013jta, Goudelis:2013uca, 
Arhrib:2013ela,Ilnicka:2015sra, Ilnicka:2015jba, Swiezewska:2012ej, Sher:2012xb,Celis:2013rcs,Belanger:2013xza, Abe:2014gua,Krawczyk:2015vka, Belanger:2015kga, Blinov:2015qva, Diaz:2015pyv}, and in the context of International Linear Collider (ILC) \cite{Moortgat-Picka:2015yla, Aoki:2013lhm, Hashemi:2015swh} in the past.  
The study in reference \cite{Arhrib:2013ela} considers the results of LUX\cite{Akerib:2013tjd}, PLANCK\cite{Ade:2015xua}, and include the updated results from LHC in references \cite{Ilnicka:2015sra, Ilnicka:2015jba},  which has provided a comprehensible analysis to provide the available parameter space regions\footnote{In the present study, bechmark points resulting from the analysis of \cite{Ilnicka:2015jba} are used.}. Most of these studies focus on the pair production of the inert scalars, and consider final states involving leptons and missing energy. The purely hadronic channels are generically marred by the large irreducible background owing to the hadronic environment of the LHC. A comprehensive report on the IDM search at Run 2 of the LHC is provided by the report of the Dark Matter Forum \cite{Abercrombie:2015wmb}. For the ILC, the effect of IDM on the triplet Higgs couplings is studied by Ref. \cite{Arhrib:2015hoa}.
 In this work, we consider the dijet along with missing energy as the signature of IDM, and explore the possible parameter reach at the LHC, with moderate to high luminosity. Apart from the pair production and subsequent cascade decays, this channel receives significant contribution from the vector boson fusion (VBF)\footnote{While this manuscript was being prepared, the study of Ref. \cite{Brooke:2016vlw} on dark matter searches through VBF appeared.} , $t-$channel with the invisible Higgs ($H$) radiating from the mixed propagator, and the $s-$channel with quartic coupling involving $H$ and $W/Z$. 
\vskip .2cm
\hskip .8cm This paper is organized in the following way. In Section II, we describe the model including the present theoretical and experimental constraints available on the model parameters. In Section III, we discuss our analysis, and finally,  in Section IV we present the summary and conclusions of our study.

\section{Inert Doublet Model}

The IDM has one additional scalar doublet (under $SU(2)_L$), compared to the SM. This additional scalar, denoted by $\Phi_2$ is odd under a discrete $Z_2$ symmetry imposed, while all the SM fields are even under this new symmetry. This $Z_2$ symmetry prohibits the Yukawa interactions of $\Phi_2$ with the SM fields.  The inert doublet, however, can have direct interaction with the gauge fields, providing the mechanism to generate the corresponding particles. A consequence of the $Z_2$ symmetry is that the lightest particle state belonging to $\Phi_2$ is stable, and thus providing a candidate dark matter. 
Denoting the SM scalar doublet as $\Phi_1$, the scalar potential  respecting $SU(2)_L$ $\otimes$ $U(1)_Y$ gauge invariance is given by

\begin{equation}
\begin{aligned}
V(\Phi_1,\Phi_2)=  \mu_1^2|\Phi_1|^2 +\mu_2^2|\Phi_2|^2+\frac{\lambda_1}{2}|\Phi_1|^4+\frac{\lambda_2}{2}|\Phi_2|^4+\lambda_3|\Phi_1|^2|\Phi_2|^2\\+\lambda_4|\Phi_1^\dag \Phi_2|^2 + \left\{\frac{\lambda_5}{2}(\Phi_1^\dag \Phi_2)^2 + H.c.\right\},
\end{aligned}
\label {c}
\end{equation}

In the CP-conserved version, the parameters $\mu_1^2$, $\mu_2^2$, $\lambda_1$, $\lambda_2$, $\lambda_3$, $\lambda_4$, $\lambda_5$ are considered to be real. In the version with exact $Z_2$ symmetry, $\Phi_2$ does not acquire any non-zero vacuum expectation value (VEV), and therefore, only the SM field, $\Phi_1$ takes part in the electroweak symmetry breaking (EWSB). After the EWSB these scalar doublets may be written in the following form in the unitary gauge.
\begin{equation}
\Phi_1=\begin{pmatrix} 0 \\  \frac{ v +h }{\sqrt 2} \end{pmatrix} , \Phi_2=\begin{pmatrix} H^+\\  \frac{H+iA}{\sqrt 2} \end{pmatrix}
\end{equation}
\hskip .1cm where  $v = 246$ GeV is the vacuum expectation value of $\Phi_1$. Apart from the SM-like Higgs $h$, this presents a neutral scalar, $H$, a neutral pseudoscalar, $A$, and two charged Higgs bosons $H^\pm$,  with the other degrees of freedom of $\Phi_1$ becoming part of the massive gauge bosons through the Higgs mechanism. 
The masses of these physical scalars can be written in terms of parameters of the potential and $v$ as 
\begin{eqnarray}
m_h^2 &=& \lambda_1 v^2\nonumber\\
m_{H^+}^2 &=& \mu_2^2 + \frac{1}{2}\lambda_3 v^2\nonumber\\
m_{H}^2 &=& \mu_2^2 + \frac{1}{2}(\lambda_3+\lambda_4+\lambda_5)v^2=m^2_{H^\pm}+
\frac{1}{2}\left(\lambda_4+\lambda_5\right)v^2\nonumber\\
m_{A}^2 &=& \mu_2^2 + \frac{1}{2}(\lambda_3+\lambda_4-\lambda_5)v^2=m^2_{H^\pm}+
\frac{1}{2}\left(\lambda_4-\lambda_5\right)v^2
\label{mass_relation}
\end{eqnarray}

We may note that the parameters are not completely free and independent of each other. There are theoretical constraints arising from the vacuum stability  \cite{Swiezewska:2012ej,Gustafsson:2010zz}, given by
\begin{equation}
 \lambda_1>0, ~~~\lambda_2> 0 , ~~~\sqrt{\lambda_1\lambda_2} + \lambda_3>0~~{\rm and}~~ \sqrt{\lambda_1\lambda_2} + \lambda_3 +\lambda_4 - |\lambda_5| > 0,
\end{equation}
and to ensure perturbativity \cite{ Barbieri:2006dq, Gustafsson:2010zz} we need to keep  $|\lambda_i| \le 8 \pi$.
Considering the case $m_{H^\pm} > (m_H,~m_A)$, Eq.~\ref{mass_relation} gives $\lambda_5<0$ for $m_H<m_A$, and $\lambda_5>0$ for $m_A<m_H$. Thus, the sign of $\lambda_5$ dictates whether $H$ or $A$ is the dark matter candidate. Apart from these theoretical constraints, we have experimental constraints coming from LEP observations \cite{Beringer:1900zz,Agashe:2014kda}. From the non-observation of $Z$ and $W$ decays to dark Higgs bosons, we require $m_H + m_A >m_Z,~~2m_{H^\pm}>m_Z$ and $m_{H,A}+m_{H^\pm}>m_W$. 
The oblique parameter, $T$, receives contributions from the IDM, which could be written in terms of the mass splittings as 
\begin{equation}
\Delta T \simeq \frac{1.08}{v^2} \big( {m_{H^\pm}-m_H}\big)\big({m_{H^\pm}-m_A} \big) = 0.07\pm0.08.
\label{eq:deltaT}
\end{equation}
SUSY searches at LEP  leads to constraints on the charged Higgs mass, $m_{H^\pm}\ge 70$ GeV \cite{Blinov:2015qva, Pierce:2007ut}, and requires $|m_A-m_H| \le 8$ GeV for $m_H\le 80$ GeV and $m_A\le 100$ GeV \cite{Ilnicka:2015sra, Ilnicka:2015jba}.
Besides these collider constraints, dark matter relic abundance and direct searches \cite{Frere:2006hp,Dolle:2009fn,Gustafsson:2012aj} put a limit on the mass of the darkmatter candidate, $40 \le m_{DM}\le 80$ GeV, where $m_{DM}$ is either $m_H$ or $m_A$ for the cases of $H$ or $A$ considered as the darkmatter, respectively. 

Coming to the LHC experiments, the bound on the invisible decay width of an SM-like Higgs boson($h$) is given to be, BR$(h\rightarrow {\rm invisible})<0.12$ \cite{Bernon:2014vta} at 95\% confidence level. This restricts the relevant coupling for $m_H\le m_h/2$. At the same time, $m_H < m_h/2$ region is ruled out considering the XENON100 and LUX measurements \cite{Goudelis:2013uca, Akerib:2013tjd, Aprile:2012nq}. Also, the future Cherenkov Telescope Array (CTA) may be able to rule out  heavier dark matter masses \cite{Queiroz:2015utg}.  Previous studies of the LHC phenomenology include Ref.~\cite{Dolle:2009ft,Miao:2010rg,Gustafsson:2012aj}. Most of these studies consider $m_{H^\pm}\le150$ GeV, for which the preferred processes are the pair productions, $H^+H^-$, $AH^\pm$ and $HH^\pm$. More recently, Ref. \cite{Belanger:2015kga} studied the constraints of IDM arising through the dilepton channels, considering two representative values of $m_{H^\pm}=85~~{\rm and}~~150$ GeV, focusing on the parameter region which are complementary to those accessible by dark matter direct searches and Higgs invisible decay channels. This include pair production of $AH,~HH,~AH$ and $H^{\pm}H^{\mp}$ decaying through to the final state of two leptons and missing energy.  

In this article, we shall focus on the dijet plus missing energy signal arising in the IDM scenario. As we shall see in the next section, this signal can originate from the production of $H^\pm$ in association with $H$, with the subsequent decay of the charged Higgs, as well as from other VBF channels, and $s-$channels with quartic $VVHH$ couplings, where $V=Z, W$, and $t-$channel with mixed propagator, radiating $HH$. 

\section{Discussion}

Discovery of the charged Higgs boson will provide a smoking gun signature of the multi-Higgs models. Compatibility of such a scenario, and further identification of the couplings would be one of the first steps in establishing a specific multi- Higgs model. The prominent production processes involving the charged Higgs bosons at the LHC are $H^+H^-,~~H^+A,~~H^+H$. In the IDM model, $H^+$ predominantly decay into $W^+H$, and $A$ decays mosly to $ZH$, leaving missing energy in all cases and making it almost impossible to reconstruct the events. 
 The magnitude of the cross section will depend on the masses and couplings of the scalars involved. The couplings are dictated by the gauge coupling, and therefore are fixed. Our interest is to explore the low, intermediate and high mass regions of $m_{H^+}$ values. For our study we have taken $\{ m_{H^+}, ~m_{A}, ~m_H, ~m_h, ~\lambda_l$, $\lambda_2$ $\}$ as our free parameter set, where masses can be expressed in terms of  the parameters available in the potential as given by  Eq.~\ref{mass_relation}, and $\lambda_l$ = $\frac{1}{2}$  ($\lambda_3$+$\lambda_4$+$\lambda_5$) is the combination  relevant to the  couplings of the dark matter candidate, $H$. Considering all the dark matter and collider constraints available presently, the following benchmark points (BP) \footnote{The BP's are provided by A.~Ilnicka and T.~Robens. We acknowledge their help, and refer the reader to their work \cite{Ilnicka:2015sra, Ilnicka:2015jba} for a detailed study of the parameter space of the model compatible with the darkmatter and collider considerations.} are selected for our study.

\noindent
{Benchmark Points used in the Table:}\\[2mm]
{BP1}: $m_{H^+}$ = 80 , $m_{A}$ = 75.4, $m_H$ = 65 , $m_h$ = 125.1 , $\lambda_l$ = 0.006 , $\lambda_2$ = 0.1 \\
{BP2}: $m_{H^+}$ = 150 , $m_{A}$ = 138.6 , $m_H$ = 65 , $m_h$ = 125.1 , $\lambda_l$ = 0.009 , $\lambda_2$ = 0.1 \\
{BP3}: $m_{H^+}$ = 200 , $m_{A}$ = 189.5, $m_H$ = 65 , $m_h$ = 125.1 , $\lambda_l$ = 0.009 , $\lambda_2$ = 0.1 \\
{BP4}: $m_{H^+}$ = 300 , $m_{A}$ = 289.3, $m_H$ = 65 , $m_h$ = 125.1 , $\lambda_l$ = 0.009 , $\lambda_2$ = 0.1 \\
{BP5}: $m_{H^+}$ = 400 , $m_{A}$ = 397.6, $m_H$ = 65 , $m_h$ = 125.1 , $\lambda_l$ = 0.009 , $\lambda_2$ = 0.1 \\
{BP6}: $m_{H^+}$ = 500 , $m_{A}$ = 494.0, $m_H$ = 65 , $m_h$ = 125.1 , $\lambda_l$ = 0.009 , $\lambda_2$ = 0.1 \\

We have fixed the mass of the dark matter candidate to be $m_H=65$ GeV, in order to avoid the invisible decay of the SM Higgs boson to a pair of DM. However, we have confirmed that, varying the mass slightly, within the window available, as described above does not bring in any significant change in our conclusions. We have then chosen different representative values of $m_{H^+}$, the main object of our study. From Eq.~\ref{eq:deltaT}, this then naturally limits the value of $m_A$ to be close to $m_{H^+}$. The values considered in the BP's are obtained from a random scan, satisfying all the dark matter and collider constraints mentioned above. 
The production cross section of the Higgs boson pairs at the LHC with $\sqrt{s}= 13$ TeV\footnote{The cross sections do not change significantly at 14 TeV LHC, and the conclusions drawn in this work are expected to be valid at this centre of mass energy as well.} for these benchmark points are given in Table~\ref{table:cs_Hpair}.  $H^+$ dominantly decay to $W^+H$ with a BR of almost 100\% for $m_{H^+}> (m_H+m_W)$. For masses below this, both $W^{+*}H$ and $W^+A^*$ could contribute. For the BP's considered here, $m_A$ is close to $m_{H^+}$ in all cases, and therefore the decay channel  $W^+A^*$ has a maximum contribution of around 2\% for $m_{H^+}=80$ GeV case (BP1).
 

\begin {table}[h!]
\begin{center}
\begingroup
\fontsize{6pt}{7pt}\selectfont
{
    \begin{tabular}{ | l | r | r | r| r | }
    \hline\hline
Benchmark&\multicolumn{4}{c|}{cross sections in fb}\\ \cline{2-5}
Points &   $p p \rightarrow H^\pm  H $  & $p p \rightarrow H^\pm A $ &   $p p \rightarrow H A $ &   $p p \rightarrow H^{+} H^{-} $  \\ \hline
BP1 & 1235.0 & 954.2 &851.8 & 446.5 \\ \hline
BP2   & 257.9& 96.7 & 179.6 & 46.4 \\ \hline
BP3 & 110.9 & 31.3 & 71.8 & 15.8\\ \hline
BP4  & 29.8 & 6.1 & 18.1 & 3.2  \\ \hline
BP5   & 10.7 & 1.7 & 5.8 & 0.9 \\ \hline
BP6   & 4.6 & 0.6 & 2.5 & 0.3  \\ \hline
\hline
  \end{tabular}
}
\endgroup
\caption{{ The production cross section for different Higgs pairs at the LHC ($\sqrt{s} = 13$ TeV) for different BPs considered.}}
\label{table:cs_Hpair}
\end{center}
\end{table}

\begin{figure}[h] \centering
\vspace{-10mm}
\begin{tabular}{c c c c}
\hspace{-5mm}
\includegraphics[angle=0,width=40mm]{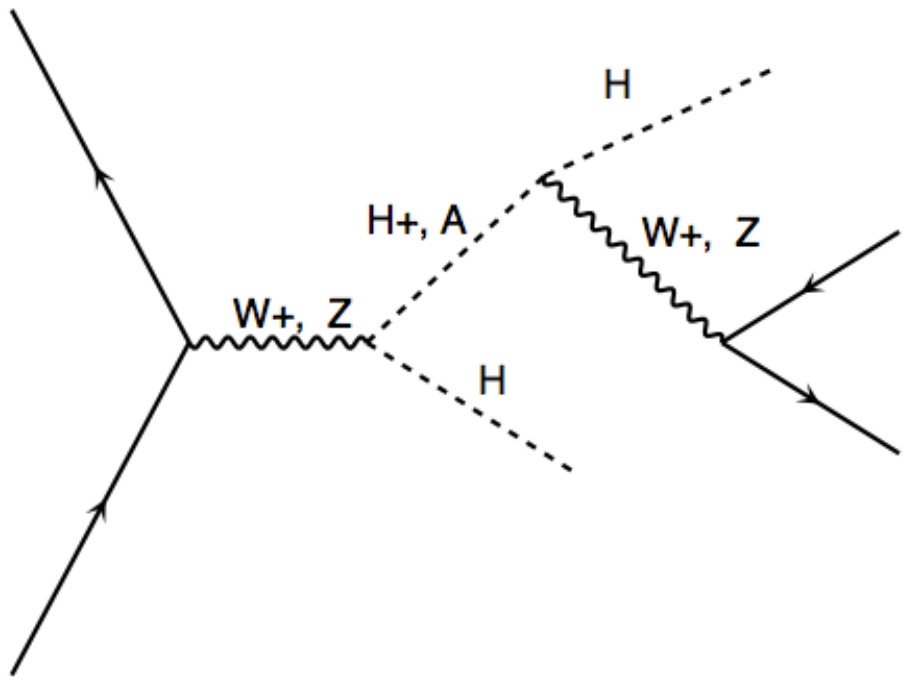}&\hspace{-10mm}
\includegraphics[angle=0,width=40mm]{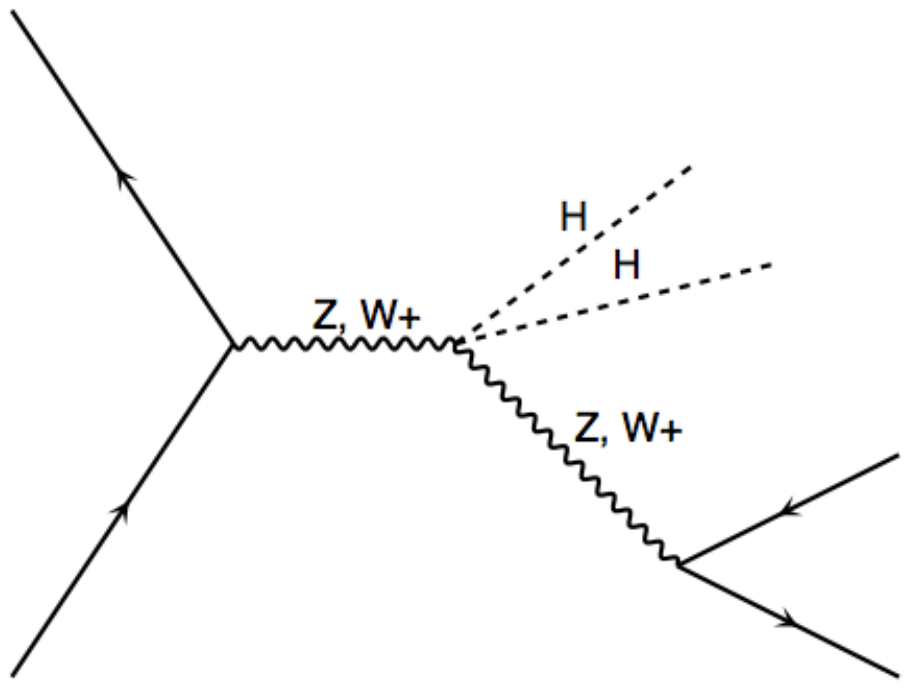}&\hspace{-10mm}
\includegraphics[angle=0,width=40mm]{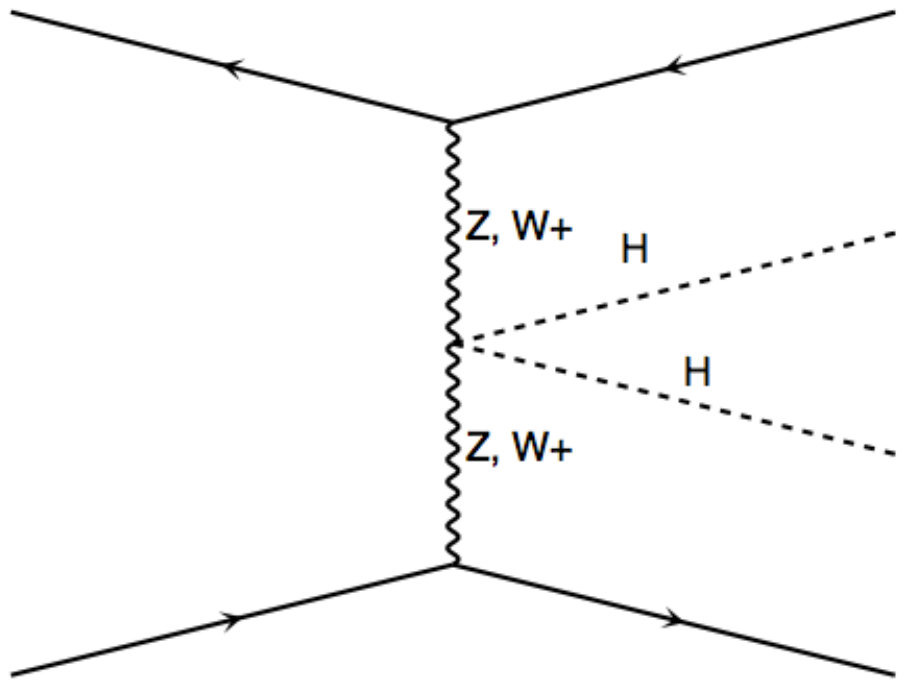}&\hspace{-10mm}
\includegraphics[angle=0,width=40mm]{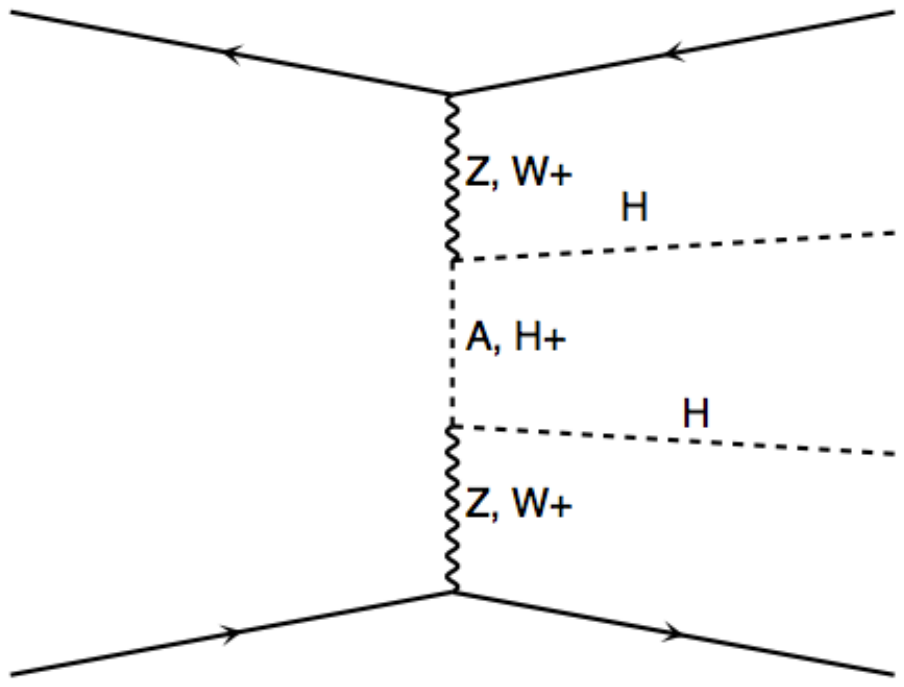}
\end{tabular}
\vspace{-15mm}
\caption {Typical Feynman diagrams from a set of such diagrams illustrate the production of $2j+HH$. }
\label{fig:feynman_diag1}
\end{figure}

\begin{figure}[h] \centering
\begin{tabular}{c c c}
\hspace{-5mm}
\includegraphics[angle=0,width=40mm]{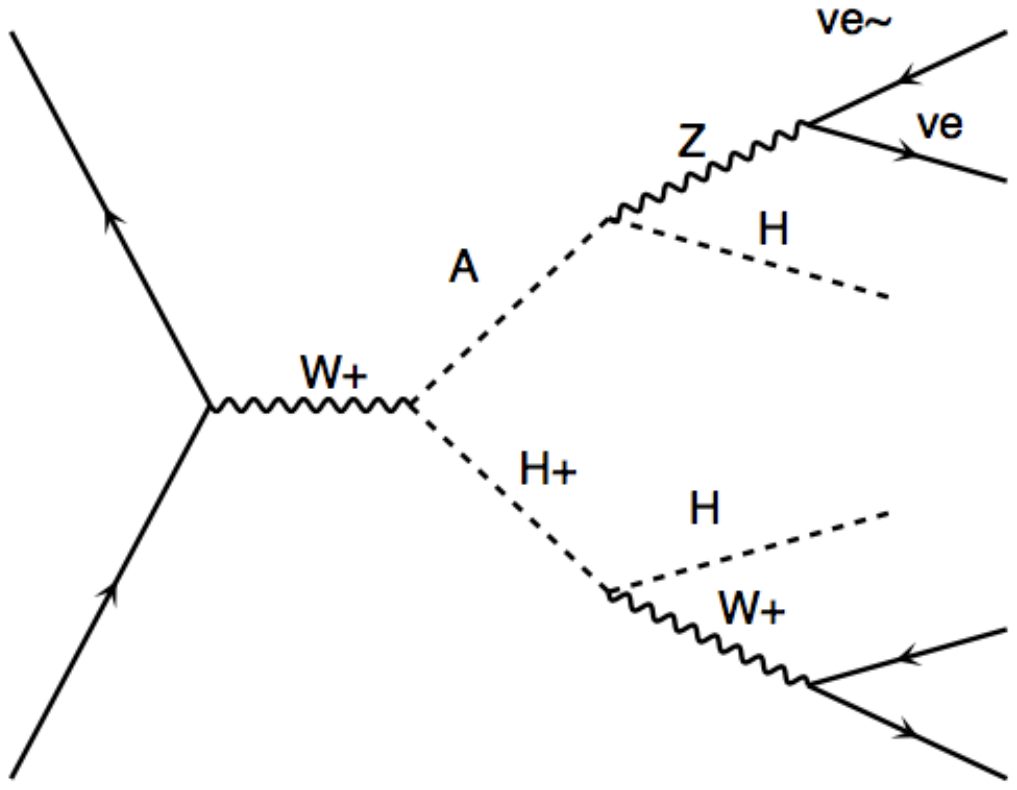}&
\includegraphics[angle=0,width=40mm]{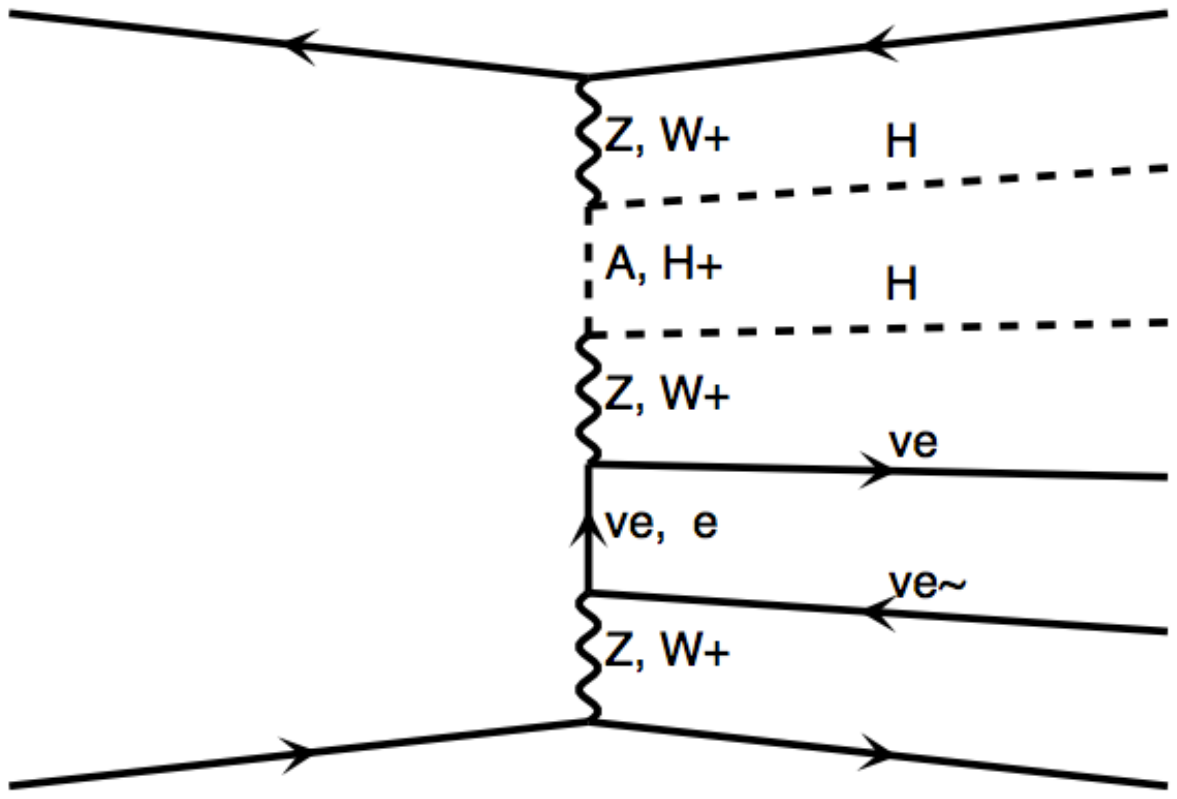}&
\includegraphics[angle=0,width=40mm]{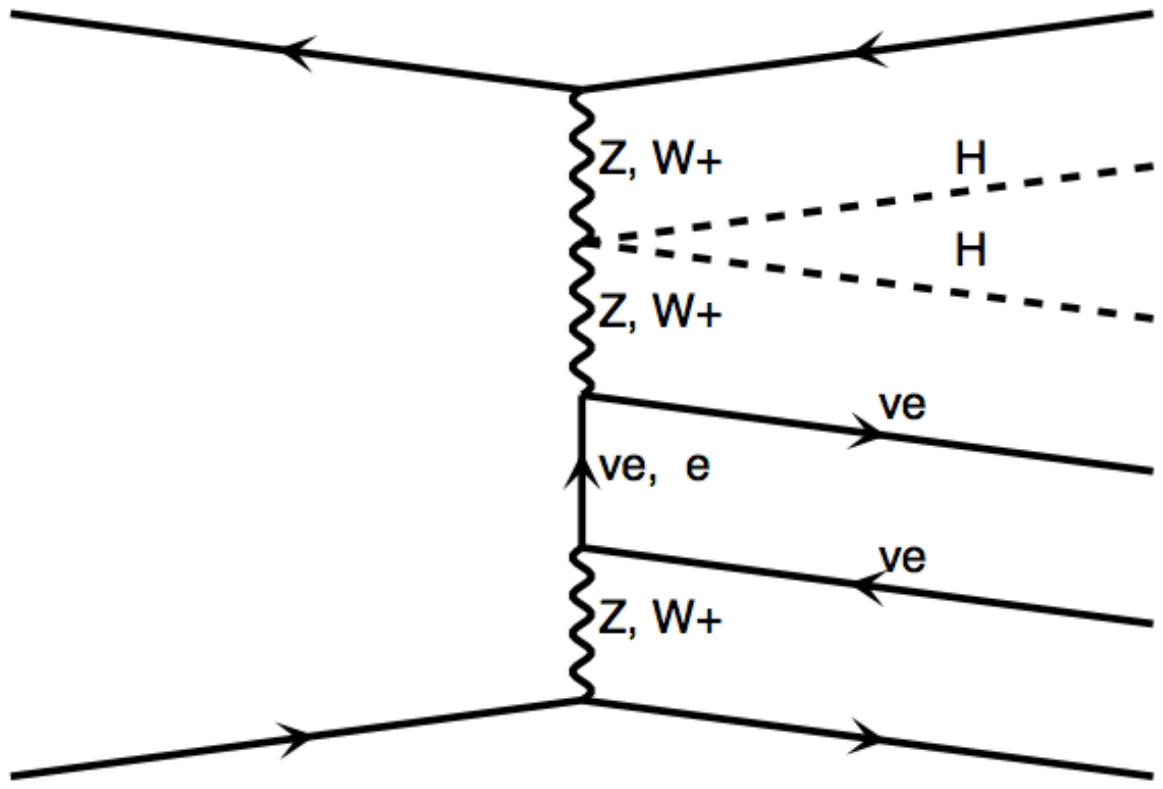}
\end{tabular}
\vspace{-15mm}
\caption {Typical Feynman diagrams from a set of such diagrams illustrate the production of $2j+\nu \bar \nu HH$. }
\label{fig:feynman_diag2}
\end{figure}

\begin {table}[h!]
\begin{center}
\begingroup
\fontsize{6pt}{7pt}\selectfont
{
    \begin{tabular}{ | l | r | r | r| r |r| r|r|}
    \hline\hline
&\multicolumn{4}{c|}{$2j+HH$ cross sections in fb}&\multicolumn{2}{c|}{$2j+\nu \bar \nu HH$ cross sections in fb}&\\ 
&\multicolumn{4}{c|}{(different channels)}&\multicolumn{2}{c|}{(different channels)}&\\ \cline{2-7}
   Benchmark &   (i) $ H^\pm  H $  & (ii): $H A $ &  total cascade &(A):$p p \rightarrow 2j+HH$ &$H^\pm A$&(B):$p p \rightarrow 2j+\nu\bar \nu HH$ &\\
   Points&&& (i)+(ii)&(all inclusive)&&(all inclusive)&(A)+(B)  \\ \hline
BP1 & 1.2 & 0.008& 1.2&1.9 &0.2&27.5& 29.4\\ \hline
BP2   & 134.5 & 36.3 & 170.8 &184.4 &9.2&17.5&201.9\\ \hline
BP3 & 53.1 & 29.0 & 82.1&86.7 &2.9&6.5&93.2\\ \hline
BP4  & 15.3 & 7.8 & 23.1 &27.5 &0.6&1.6&29.1\\ \hline
BP5   & 5.5 & 2.5 & 8.0&13.9 &0.2&0.5&14.4\\ \hline
BP6   & 2.3 & 1.1 & 3.4 &10.5  &0.005&0.2&10.7\\ \hline

\hline
  \end{tabular}
}
\endgroup
\caption{{ Cross section(in fb) at $\sqrt s$ = 13 TeV for specific benchmark points, showing the significance of the VBF and $t-$channel contributions.}}
\label{table:cs_compare}
\end{center}
\end{table}
The detector-level final states, considering the decay of $W$ and $Z$, are (i) the purely hadronic with jets and missing energy, (ii) jets and leptons with missing energy, and (iii) purely leptonic with missing energy. 
$H^+H^-$ thus leads to the final states with $4j+MET$, $2j+l+MET$ and $2l+MET$, with the $W$ pair from the $H^{\pm}$ decay hadronically, semileptonically, and leptonically, respectively. The $H^+A$, similarly, leads to $4j+MET$, $2j+2l+MET$, $2j+l+MET$,  $2j+MET$, $3l+MET$ and $l+MET$, with $W$ and $Z$ decaying into hadronic jets and leptons, as is the case may be. The other two pairs, $H^+H$ and $AH$ give the final states, $2j+MET$, $2l+MET$, and $l+MET$. The $4j+MET$ final state signal is almost impossible to ressurect from the huge hadronic backgrounds at the LHC. Possibilities at the 100 TeV proton-proton collider could be promising. This is being explored in an independent anlaysis, and we defer the discussion to a separate publication in preparation\cite{4jmet}. The leponic final states are studied in the literature in the context of LHC. In this article, we shall focus on the dijet plus missing energy signal at the LHC. 
The $2j+MET$ arises through the cascade decays, $H^+A \rightarrow (W^+H)~(ZH)\rightarrow (jjH)~(\nu\bar \nu H)$, and $H^+H \rightarrow (W^+H)~H\rightarrow (jjH)~H$. 
Apart from these cascade decays, this final state could arise from the VBF to $HH$, and the $t-$channel with mixed propagator involving $W(Z)$ and $H^+(A)$, as shown in the illustrative example Feynman diagrams in Fig.~\ref{fig:feynman_diag1} and \ref{fig:feynman_diag2}


The contributions arising from the VBF and $s-$channel with quartic couplings, and the $t-$channel processes could be significant, depending on the mass ranges of the Higgs bosons considered. In Table~\ref{table:cs_compare}, we compare the cross sections from cascade decays,  separately, along with the total $2j+MET$ cross section, including any possible interference effects. 
The cross sections are obtained through Madgraph5, and after employing the basic cuts on transverse momentum, $p_T(j)>20$ GeV and pseudo rapidity, $|\eta_j|<5.0$ to the two jets, and demanding a separation between the jets of $\Delta R_{j_1j_2} > 0.4$. The cross sections quoted in Column 4 compared with that in Column 5 reveals the significance of contributions coming from other-than-cascade-decay processes in the case of $2j+HH$ final state, and the possible interference of these two categories. The trend is clear, that the contributions from cascade decay goes down drastically, as $m_{H^+}$ increases, whereas, the other contributions seem to remain steady within the range of 5 to 7 fb. The slightly different behaviour at $m_{H^+}=150$ GeV is possibly due to the very large and dominating contribution from the cascade-decay channel, where even the interference effects could play a significant role. The case of $2j+\nu\bar \nu+HH$ on the other hand gives a slightly different picture. Beyond $m_{H^+}=300$ GeV, the contributions are very small. On the contrary, it contributes significantly at lower $m_{H^+}$ values. Looking at the $t$-channel topology of the additional contributions, it is likely that the two jets are produced with large $p_T$, and therefore the cross section is not reduced by removing the soft jets, as is employed in getting the cross sections in Table~\ref{table:cs_compare}. 
Background to the process $p p \rightarrow  2j+MET$ in IDM arises  through the SM processes of $2j+\nu\bar \nu$ and $W+2j$, where the latter contribute in the leptonic decay of $W$ with soft leptons, or leptons missing into the beam pipe. The cross section for these background processes at the 13 TeV LHC are 955 pb and 51 pb, respectively.
In the following we shall discuss the signal and background, including the kinematic distributions, and establish the reach of the LHC in probing IDM through this channel.

\begin{figure}[h] \centering
\begin{tabular}{c c}
\hspace{-5mm}
\includegraphics[angle=0,width=75mm]{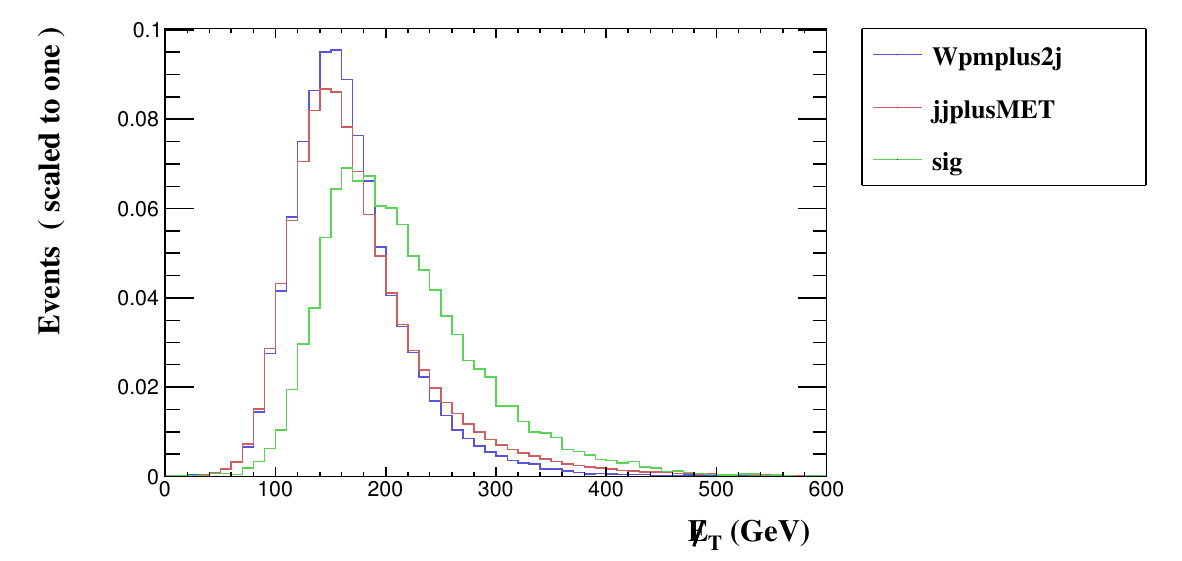}
\includegraphics[angle=0,width=75mm]{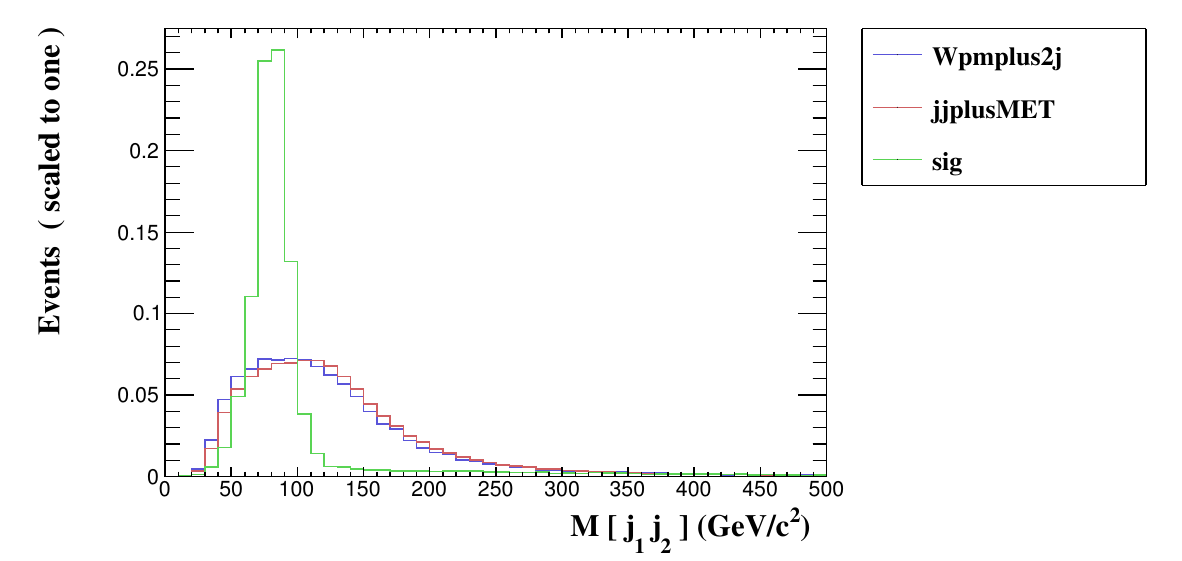}\\
\includegraphics[angle=0,width=75mm]{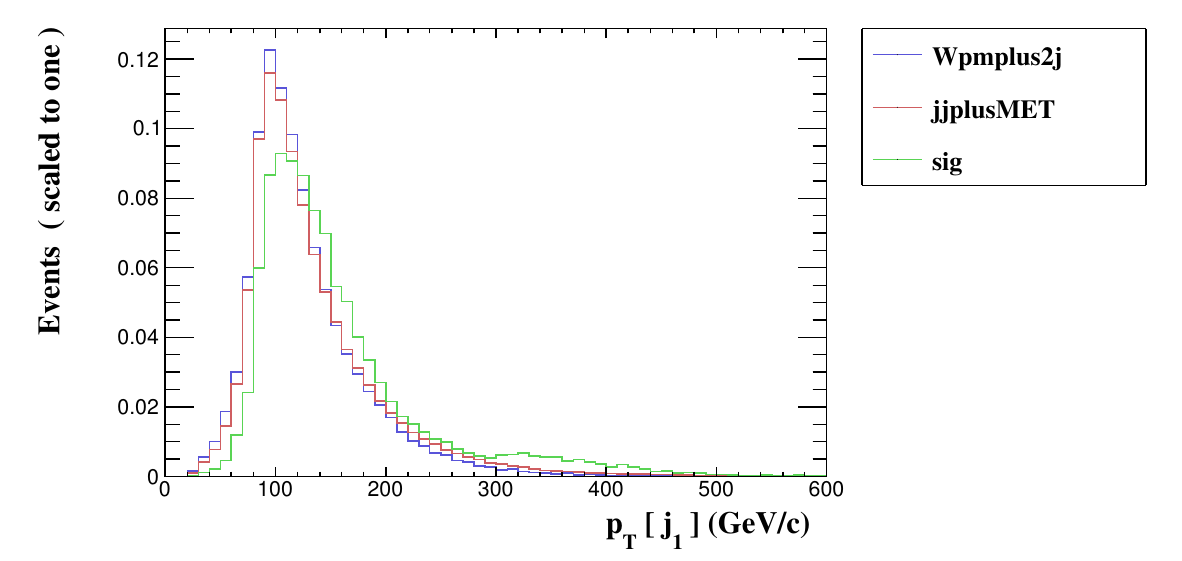}
\includegraphics[angle=0,width=75mm]{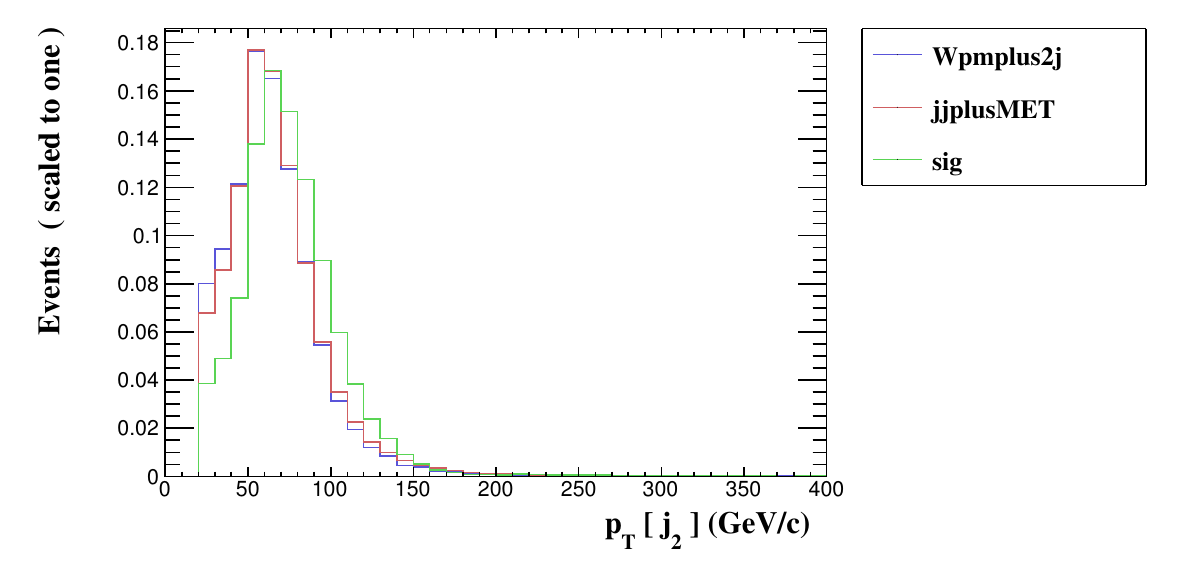}
\end{tabular}
\vspace{-1mm}
\caption {Kinematic distributions of the $2j+MET$ events for the scenario of BP4 corresponding to 
$m_{H^+}=300$ GeV at 13 TeV LHC, after applying the basic selection criteria as discussed in the text.}
\label{fig:dist1_cuts}
\end{figure}

We have used MADGRAPH5 \cite{Alwall:2014hca} for our analysis with the IDM model imported through the UFO generated from the publicly available FeynRules \cite{feynrules} interface. The signal and background processes are generated through MADGRAPH5 along with basic acceptance cuts employed.  For hadronization, we have considered PYTHIA6\cite{Sjostrand:2006za} inside MADGRAPH5 with the options of ISR and FSR included. 
A study of the $p_T({jets})$ of the signal concluded that we can employ the basic cuts of $p_T({j_1})>80$ GeV, and $p_T({j_2})>50$ GeV, $|\eta_{jets}|<5.0$, and a jet separation of $0.4<\Delta R_{j_1j_2}<2.0$, at the generation level, without compromising the signal events significantly. This reduces the effective fiducial cross sections of the signal to 1.3 fb, 14 fb, 16 fb, 9.8 fb, 4.8 fb and 2.5 fb, for the cases of BP1, BP2, BP3, BP4, BP5 and BP6, respectively. We generated 50000 signal events in all cases. 
The background cross sections are reduced to 12.94 pb and 0.405 pb for $jj+\nu\bar \nu$ and $Wjj$, respectively. We generated 
1300000, and 100000 events for these two backgrounds respectively, which provides sufficient statistics at 100 fb$^{-1}$ luminosity.
  The events thus generated are then analysed with the help of MADANALYSIS5 \cite{Conte:2012fm}, using the inbuilt interface with Fastjet and DELPHES with the CMS card. For  jet reconstruction with Fastjet, we used anti-$k_t$ algorithm with $\Delta R=0.5$.
 The following selection cuts are applied to optimise the signal over the background.
 The events are cleaned from soft-jets and leptons that could arise in the detector simulation, by removing the jets softer than $p_T<20$ GeV, and $|\eta|>5.0$. Events with two jets are then selected ($N(j)=2$), and also demanded that the events do not contain $b$-jets ($N(b)=0$) or leptons ($N(l)=0$). 
  In Fig.~\ref{fig:dist1_cuts}, we present some of the kinematic distributions corresponding to the case of $m_{H^+}=300$ GeV, after employing the above selection. The other BP's have similar distributions, which are not presented here.
 Learning from the distributions, we employ further selection cuts on the kinematic distributions with the aim of improving the signal significance. A set of final selection cuts, with transverse momenta of the jets, {{  $p_T({j_1})>120$ GeV and $p_T({j_2})>90$ GeV,  jet separation of $\Delta R_{j_1j_2}<1.8$, missing transverse energy  $MET > 260$ GeV, and the invariant mass of the two jets, $75 < M_{j_1j_2}<90$ GeV are considered.

With the above selection criteria, the case with low $m_{H^+}=80$ GeV of BP1 is very difficult to probe due to very small cross section available after the basic cuts employed mentioned above. Considering other BP's,
we are left with 90 signal events ($S$) for BP2 corresponding to $m_{H^+}=150$ GeV, over a background ($B$) of 8500 SM events at 1000 fb$^{-1}$ integrated luminosity. This means a significance of $\frac{S}{\sqrt{S+B}}=0.97$, which is improved to $1.68$ with 3000 fb$^{-1}$ luminosity. The signal events are about 1\% of the background events, and it requires a very controlled systematics to see the events. However, when we move on to larger $m_{H^+}$ values, situation gets better. At BP3, BP4 and BP5 with $m_{H^+}=200, ~300$ and the 400 GeV, respectively, the signal events are 198,  168 and 120 with a luminosity of 1000 fb$^{-1}$.  This corresponds to a signal significance of 2.12, 1.80 and 1.29, which are improved to 3.68, 3.13 and 2.24, respectively, at 3000 fb$^{-1}$ luminosity. The ratio of the signal to background events is now a somewhat better 2.32\%,  1.97\% and 1.41, for the respective cases. The other  BP with $m_{H^+}= 500$ GeV (BP6) does not spare that well with these selection criteria. The number of signal events at 1000 fb$^{-1}$ corresponding to this BP is 70, giving a significance of 0.76, which is improved to 1.31 at 3000 fb$^{-1}$. The signal to background ratio is now 0.82\%, which is less than the expected systematic uncertainty. We have summarised the above results in Table~\ref{table:selection}. We have employed a uniform selection criteria for all the BP's considered, keeping in mind that such analysis will be easier from the point of view of data analysis.   We understand that, the systematic uncertainties could play a critical role while looking for BSM effects with such large SM background events expected. While we do not attempt an involved analysis including the effects of the systematic uncertainties, we have looked at the effects on the significance with an assumed uncertainty of 1\% on the background, and 10\% systematic uncertainty on the signal events. The resulting significance computed using the formula $\frac{S}{\sqrt{B+(0.01\times B)^2+(0.1\times S)^2}}$ is presented in the Table ~\ref{table:selection}. Clearly, the BP3 leaves a significance  of about 2, which is sufficient to give a clear hint of a possible BSM signal. The significance corresponding to BP4 and BP5 lie between 1 and 2, while the other two BP's (BP2 and BP6) provide significance less than one. 
}}

\begin {table}[h!]
\begin{center}
\begingroup
{\small
    \begin{tabular}{|l|l|r|r|r|r|r|r|r|r|}
    \hline\hline
Cuts employed&$B$&BP's&$S$&$\frac{S}{B}$ \%& \multicolumn{2}{c|}{$\frac{S}{\sqrt{S+B}}$}& \multicolumn{2}{c|}{\small{$\frac{S}{\sqrt{B+(0.01\times B)^2+(0.1\times S)^2}}$}}\\  \cline{6-9}
&&&&&{\small{1 /ab}}&{\small{3 /ab}}&{\small{1 /ab}}&{\small{3 /ab}}\\
\hline
$N(j)=2$, $N(b)=0$, $N(l)=0$,&&BP2&90&1.05&0.97&1.68&0.72&0.89\\
 \cline{3-9}
$MET>260$ GeV, &8500&BP3&198&2.32&2.12&3.68&1.56&1.94\\\cline{3-9}
$p_T(j_1)>120$ ,  $p_T(j_2)>90$,&&BP4&168&1.97&1.80&3.13&1.33&1.65\\  \cline{3-9}
$75\le M_{j_1j_2}\le 90$ GeV,&&BP5&120&1.41&1.29&2.24&0.95&1.19\\  \cline{3-9}
 $\Delta R_{j_1j_2}<1.8$&&BP6&70&0.82&0.76&1.31&0.56&0.70\\ 
\hline
  \end{tabular}
}
\endgroup
\caption{Generic selection cuts employed to optimise the $S/B$ ratio and the signal significance at a 13 TeV LHC, along with the number of signal events $(S)$, number of background events $(B)$ corresponding to the different Benchmark Points (BP's) considered at integrated luminosity of 1000 fb$^{-1}$.  Significances corresponding to a luminosity of 3000 fb$^{-1}$ are also quoted.  Significance with assumed systematic uncertainties are given in the last two columns.}
\label{table:selection}
\end{center}
\end{table}

Please note that the above analysis is performed, keeping in mind a generic set of selection criteria that could be employed while searching for signals of the BSM scenarios, the presence of IDM in the present case. We conclude that, in contrast to the phenomenological studies involving leptonic final states, our analysis present a way to probe the large $m_{H^+}$ regions up to a value of around 300 - 400  GeV  with high, but achievable, luminosity at the LHC through the $dijet+MET$ channel. Beyond these masses, establishing signals above background is somewhat difficult. However, upto even 500 GeV mass ranges, it is possible to probe the model with somewhat smaller significance.  

\section{Conclusion}

The inert doublet model presents an interesting scenario within the multi-Higgs models, with a candidate dark matter, resulting in distinct phenomenology compared to other models like the 2HDM and MSSM. The model is compatible with all the experimental constraints arising from dark matter searches, as well as from collider experiments including the recent LHC measurements. In a specific scenario, we have considered the mass hierarchy of $m_{H^+}>m_A>m_H$, so that the neutral scalar is the dark matter candidate. 
We have considered the possibility to probe the model through $2j+MET$ signal at the LHC with high luminosity. This signal arises in IDM through the cascade decay of pair production of Higgs bosons of the dark sector, along with other production mechanism like VBF, $s-$channel with quartic Higgs-gauge couplings, $t-$channel with two $H$ radiating from the gauge-Higgs mixed propagator. 
Contributions of cascade alone are significantly reduced at larger $m_{H^+}$ values, whereas the contributions from other channels are somewhat independent of the Higgs mass, and remains at a few fb level throughout. This provides a promising possibility to probe scenarios with $m_{H^+}>150$ GeV, which is almost impossible with other channels studied in the literature \cite{Dolle:2009ft,Miao:2010rg,Gustafsson:2012aj}.

We have specifically considered a few benchmark points with $m_{H^+}$ ranging from 80 GeV to 500 GeV. The effect of systematic effects are included through an assumed 1\%  and 10\% uncertainties on the background and signal events. {{The best case scenarios are the cases with $m_{H^+}$ around 200 - 400 GeV, which could be probed at the LHC with about 3000 fb$^{-1}$ integrated luminosity with a signal significance of about 2 for $m_{H^+}= 200$ GeV, and slightly lower, but still better than one for the larger mass regions.   For higher mass case of $m_{H^+}= 500$ GeV, the significance is smaller than one, and cases with $m_{H^+}$ beyond this range are harder to probe even at such high luminosity.}} The low mass scenarios with $m_{H^+}=80$ GeV is also very difficult, mainly owing to the fact that the jets arising from these are too soft, and hard to isolate from the QCD background. 

In summary, it is clear that probing $2j+MET$ provides good handle on the search for IDM at LHC, and complements search through other leptonic channels. For scenarios like intermediate range of charged Higgs mass, this channel adds to other searches through leptonic and semi-leptonic channels. For larger mass range, where the leptonic channels become inefficient, the dijet plus missing energy channel discussed here proves to be an effective probe mechanism, albeit with the need of large luminosity of the order of 3000 fb$^{-1}$.

\vskip 10mm
\noindent
{\bf \large Acknowledgements:} We thank A.~Ilnicka and T.~Robens for providing the benchmark points. We thank Benjamin Fuks for his prompt and constant help in resolving issue with Madanalysis5, and Olivier Mattelaer for help with Madgraph. Also, we would like to thank Subhadeep Mondal, Jyotiranjan Beuria, Nabarun Chakrabarty for help with Madgraph and Madanalysis. We thank Genevieve Belanger for useful discussions. Work of PP and SS is supported by the SERB, DST, India research grant, EMR/2015/000333, and FIST-DST, India grant, SR/FST/PS-II-020/2009 for computing facilities.


\begin{thebibliography}{100}








\bibitem{Aad:2012tfa} 
  G.~Aad {\it et al.}  [ATLAS Collaboration],
  Phys.\ Lett.\ B {\bf 716}, 1 (2012)
  [arXiv:1207.7214 [hep-ex]].

\bibitem{Chatrchyan:2012ufa} 
  S.~Chatrchyan {\it et al.}  [CMS Collaboration],
  Phys.\ Lett.\ B {\bf 716}, 30 (2012)
  [arXiv:1207.7235 [hep-ex]].


\bibitem{Deshpande:1977rw} 
  N.~G.~Deshpande and E.~Ma,
  Phys.\ Rev.\ D {\bf 18}, 2574 (1978).

\bibitem{Ma:2006km} 
  E.~Ma,
  Phys.\ Rev.\ D {\bf 73}, 077301 (2006)
  [hep-ph/0601225].
  
\bibitem{Ma:2006fn} 
  E.~Ma,
  Mod.\ Phys.\ Lett.\ A {\bf 21}, 1777 (2006)
  doi:10.1142/S0217732306021141
  [hep-ph/0605180].

\bibitem{Krauss:2002px} 
  L.~M.~Krauss, S.~Nasri and M.~Trodden,
  Phys.\ Rev.\ D {\bf 67}, 085002 (2003)
  doi:10.1103/PhysRevD.67.085002
  [hep-ph/0210389].

\bibitem{Kubo:2006yx} 
  J.~Kubo, E.~Ma and D.~Suematsu,
  Phys.\ Lett.\ B {\bf 642}, 18 (2006)
  doi:10.1016/j.physletb.2006.08.085
  [hep-ph/0604114].

\bibitem{Sierra:2008wj} 
  D.~Aristizabal Sierra, J.~Kubo, D.~Restrepo, D.~Suematsu and O.~Zapata,
  Phys.\ Rev.\ D {\bf 79}, 013011 (2009)
  doi:10.1103/PhysRevD.79.013011
  [arXiv:0808.3340 [hep-ph]].

\bibitem{Suematsu:2009ww} 
  D.~Suematsu, T.~Toma and T.~Yoshida,
  Phys.\ Rev.\ D {\bf 79}, 093004 (2009)
  doi:10.1103/PhysRevD.79.093004
  [arXiv:0903.0287 [hep-ph]].

\bibitem{Suematsu:2010gv} 
  D.~Suematsu, T.~Toma and T.~Yoshida,
  Phys.\ Rev.\ D {\bf 82}, 013012 (2010)
  doi:10.1103/PhysRevD.82.013012
  [arXiv:1002.3225 [hep-ph]].

\bibitem{Suematsu:2011va} 
  D.~Suematsu,
  Eur.\ Phys.\ J.\ C {\bf 72}, 1951 (2012)
  doi:10.1140/epjc/s10052-012-1951-z
  [arXiv:1103.0857 [hep-ph]].

\bibitem{Kashiwase:2012xd} 
  S.~Kashiwase and D.~Suematsu,
  Phys.\ Rev.\ D {\bf 86}, 053001 (2012)
  doi:10.1103/PhysRevD.86.053001
  [arXiv:1207.2594 [hep-ph]].

\bibitem{Kashiwase:2013uy} 
  S.~Kashiwase and D.~Suematsu,
  Eur.\ Phys.\ J.\ C {\bf 73}, 2484 (2013)
  doi:10.1140/epjc/s10052-013-2484-9
  [arXiv:1301.2087 [hep-ph]].

\bibitem{Chakrabarty:2015yia} 
  N.~Chakrabarty, D.~K.~Ghosh, B.~Mukhopadhyaya and I.~Saha,
  Phys.\ Rev.\ D {\bf 92}, no. 1, 015002 (2015)
  doi:10.1103/PhysRevD.92.015002
  [arXiv:1501.03700 [hep-ph]].


\bibitem{Barbieri:2000gf} 
  R.~Barbieri and A.~Strumia,
  hep-ph/0007265.


\bibitem{Barbieri:2006dq} 
  R.~Barbieri, L.~J.~Hall and V.~S.~Rychkov,
  Phys.\ Rev.\ D {\bf 74}, 015007 (2006)
  doi:10.1103/PhysRevD.74.015007
  [hep-ph/0603188].

\bibitem{Casas:2006bd} 
  J.~A.~Casas, J.~R.~Espinosa and I.~Hidalgo,
  Nucl.\ Phys.\ B {\bf 777}, 226 (2007)
  doi:10.1016/j.nuclphysb.2007.04.024
  [hep-ph/0607279].


\bibitem{Hambye:2007vf} 
  T.~Hambye and M.~H.~G.~Tytgat,
  Phys.\ Lett.\ B {\bf 659}, 651 (2008)
  doi:10.1016/j.physletb.2007.11.069
  [arXiv:0707.0633 [hep-ph]].


    
\bibitem{Cao:2007rm} 
  Q.~H.~Cao, E.~Ma and G.~Rajasekaran,
  Phys.\ Rev.\ D {\bf 76}, 095011 (2007)
  doi:10.1103/PhysRevD.76.095011
  [arXiv:0708.2939 [hep-ph]].
  
\bibitem{Dolle:2009ft} 
  E.~Dolle, X.~Miao, S.~Su and B.~Thomas,
  Phys.\ Rev.\ D {\bf 81}, 035003 (2010)
  doi:10.1103/PhysRevD.81.035003
  [arXiv:0909.3094 [hep-ph]].

  
\bibitem{Arhrib:2012ia} 
  A.~Arhrib, R.~Benbrik and N.~Gaur,
  Phys.\ Rev.\ D {\bf 85}, 095021 (2012)
  doi:10.1103/PhysRevD.85.095021
  [arXiv:1201.2644 [hep-ph]].
  
\bibitem{Krawczyk:2013wya} 
  M.~Krawczyk, D.~Sokolowska and B.~Swiezewska,
  arXiv:1304.7757 [hep-ph].
  
\bibitem{Krawczyk:2013jta} 
  M.~Krawczyk, D.~Sokolowska, P.~Swaczyna and B.~Swiezewska,
  JHEP {\bf 1309}, 055 (2013)
  doi:10.1007/JHEP09(2013)055
  [arXiv:1305.6266 [hep-ph]].
  
\bibitem{Goudelis:2013uca} 
  A.~Goudelis, B.~Herrmann and O.~St\"al,
  JHEP {\bf 1309}, 106 (2013)
  doi:10.1007/JHEP09(2013)106
  [arXiv:1303.3010 [hep-ph]].

\bibitem{Arhrib:2013ela} 
  A.~Arhrib, Y.~L.~S.~Tsai, Q.~Yuan and T.~C.~Yuan,
  JCAP {\bf 1406}, 030 (2014)
  doi:10.1088/1475-7516/2014/06/030
  [arXiv:1310.0358 [hep-ph]].
  
  \bibitem{Ilnicka:2015sra} 
  A.~Ilnicka, M.~Krawczyk and T.~Robens,
  arXiv:1505.04734 [hep-ph].
  
\bibitem{Ilnicka:2015jba} 
  A.~Ilnicka, M.~Krawczyk and T.~Robens,
  Phys.\ Rev.\ D {\bf 93}, no. 5, 055026 (2016)
  doi:10.1103/PhysRevD.93.055026
  [arXiv:1508.01671 [hep-ph]].
 
 

\bibitem{Miao:2010rg} 
  X.~Miao, S.~Su and B.~Thomas,
  Phys.\ Rev.\ D {\bf 82}, 035009 (2010)
  doi:10.1103/PhysRevD.82.035009
  [arXiv:1005.0090 [hep-ph]].
  
\bibitem{Gustafsson:2012aj} 
  M.~Gustafsson, S.~Rydbeck, L.~Lopez-Honorez and E.~Lundstrom,
  Phys.\ Rev.\ D {\bf 86}, 075019 (2012)
  doi:10.1103/PhysRevD.86.075019
  [arXiv:1206.6316 [hep-ph]].
  
\bibitem{Swiezewska:2012ej} 
  B.~\'SwieÅ¼ewska,
  Phys.\ Rev.\ D {\bf 88}, no. 5, 055027 (2013)
  Erratum: [Phys.\ Rev.\ D {\bf 88}, no. 11, 119903 (2013)]
  doi:10.1103/PhysRevD.88.055027, 10.1103/PhysRevD.88.119903
  [arXiv:1209.5725 [hep-ph]].
 
\bibitem{Sher:2012xb} 
  M.~Sher,
  PoS CHARGED {\bf 2012}, 015 (2012)
  [arXiv:1212.0789 [hep-ph]].

\bibitem{Celis:2013rcs} 
  A.~Celis, V.~Ilisie and A.~Pich,
  JHEP {\bf 1307}, 053 (2013)
  doi:10.1007/JHEP07(2013)053
  [arXiv:1302.4022 [hep-ph]].



\bibitem{Belanger:2013xza} 
  G.~Belanger, B.~Dumont, U.~Ellwanger, J.~F.~Gunion and S.~Kraml,
  Phys.\ Rev.\ D {\bf 88}, 075008 (2013)
  doi:10.1103/PhysRevD.88.075008
  [arXiv:1306.2941 [hep-ph]].


\bibitem{Abe:2014gua} 
  T.~Abe, R.~Kitano and R.~Sato,
  Phys.\ Rev.\ D {\bf 91}, no. 9, 095004 (2015)
  doi:10.1103/PhysRevD.91.095004
  [arXiv:1411.1335 [hep-ph]].

\bibitem{Krawczyk:2015vka} 
  M.~Krawczyk, M.~Matej, D.~Soko\'lowska and B.~ \'Swie\'zewska,
  Acta Phys.\ Polon.\ B {\bf 46}, no. 1, 169 (2015)
  doi:10.5506/APhysPolB.46.169
  [arXiv:1501.04529 [hep-ph]].


  \bibitem{Belanger:2015kga} 
  G.~Belanger, B.~Dumont, A.~Goudelis, B.~Herrmann, S.~Kraml and D.~Sengupta,
  Phys.\ Rev.\ D {\bf 91}, no. 11, 115011 (2015)
  doi:10.1103/PhysRevD.91.115011
  [arXiv:1503.07367 [hep-ph]].
  
\bibitem{Blinov:2015qva} 
  N.~Blinov, J.~Kozaczuk, D.~E.~Morrissey and A.~de la Puente,
  Phys.\ Rev.\ D {\bf 93}, no. 3, 035020 (2016)
  doi:10.1103/PhysRevD.93.035020
  [arXiv:1510.08069 [hep-ph]].

\bibitem{Diaz:2015pyv} 
  M.~A.~Díaz, B.~Koch and S.~Urrutia-Quiroga,
  Adv.\ High Energy Phys.\  {\bf 2016}, 8278375 (2016)
  doi:10.1155/2016/8278375
  [arXiv:1511.04429 [hep-ph]].
  

\bibitem{Moortgat-Picka:2015yla} 
  G.~Moortgat-Pick {\it et al.},
  Eur.\ Phys.\ J.\ C {\bf 75}, no. 8, 371 (2015)
  doi:10.1140/epjc/s10052-015-3511-9
  [arXiv:1504.01726 [hep-ph]].

\bibitem{Aoki:2013lhm} 
  M.~Aoki, S.~Kanemura and H.~Yokoya,
  Phys.\ Lett.\ B {\bf 725}, 302 (2013)
  doi:10.1016/j.physletb.2013.07.011
  [arXiv:1303.6191 [hep-ph]].
  
\bibitem{Hashemi:2015swh} 
  M.~Hashemi, M.~Krawczyk, S.~Najjari and A.~F.~\'Zarnecki,
  doi:10.1007/JHEP02(2016)187
  [arXiv:1512.01175 [hep-ph]].

%
%

   \bibitem{Akerib:2013tjd} 
  D.~S.~Akerib {\it et al.} [LUX Collaboration],
  Phys.\ Rev.\ Lett.\  {\bf 112}, 091303 (2014)
  doi:10.1103/PhysRevLett.112.091303
  [arXiv:1310.8214 [astro-ph.CO]].
  
\bibitem{Ade:2015xua} 
  P.~A.~R.~Ade {\it et al.} [Planck Collaboration],
  arXiv:1502.01589 [astro-ph.CO].

    
  
%
%
\bibitem{Abercrombie:2015wmb} 
  D.~Abercrombie {\it et al.},
  arXiv:1507.00966 [hep-ex].
  
  
%
%
\bibitem{Arhrib:2015hoa} 
  A.~Arhrib, R.~Benbrik, J.~El Falaki and A.~Jueid,
  JHEP {\bf 1512}, 007 (2015)
  doi:10.1007/JHEP12(2015)007
  [arXiv:1507.03630 [hep-ph]].
  

  
  %
\bibitem{Brooke:2016vlw} 
  J.~Brooke, M.~R.~Buckley, P.~Dunne, B.~Penning, J.~Tamanas and M.~Zgubic,
  Phys.\ Rev.\ D {\bf 93}, no. 11, 113013 (2016)
  doi:10.1103/PhysRevD.93.113013
  [arXiv:1603.07739 [hep-ph]].


\bibitem{Gustafsson:2010zz} 
  M.~Gustafsson,
  PoS CHARGED {\bf 2010}, 030 (2010)
  [arXiv:1106.1719 [hep-ph]].
  
\bibitem{Beringer:1900zz} 
  J.~Beringer {\it et al.}  [Particle Data Group Collaboration],
  Phys.\ Rev.\ D {\bf 86}, 010001 (2012).
  
    \bibitem{Agashe:2014kda} 
  K.~A.~Olive {\it et al.} [Particle Data Group Collaboration],
  Chin.\ Phys.\ C {\bf 38}, 090001 (2014).
  doi:10.1088/1674-1137/38/9/090001
  
\bibitem{Pierce:2007ut} 
  A.~Pierce and J.~Thaler,
  JHEP {\bf 0708}, 026 (2007)
  doi:10.1088/1126-6708/2007/08/026
  [hep-ph/0703056 [HEP-PH]].


\bibitem{Frere:2006hp} 
  J.-M.~Frere, F.-S.~Ling, L.~Lopez Honorez, E.~Nezri, Q.~Swillens and G.~Vertongen,
  Phys.\ Rev.\ D {\bf 75}, 085017 (2007)
  [hep-ph/0610240].

\bibitem{Dolle:2009fn} 
  E.~M.~Dolle and S.~Su,
  Phys.\ Rev.\ D {\bf 80}, 055012 (2009)
  [arXiv:0906.1609 [hep-ph]].

  
    \bibitem{Bernon:2014vta} 
  J.~Bernon, B.~Dumont and S.~Kraml,
  Phys.\ Rev.\ D {\bf 90}, 071301 (2014)
  doi:10.1103/PhysRevD.90.071301
  [arXiv:1409.1588 [hep-ph]].


\bibitem{Aprile:2012nq} 
  E.~Aprile {\it et al.} [XENON100 Collaboration],
  Phys.\ Rev.\ Lett.\  {\bf 109}, 181301 (2012)
  doi:10.1103/PhysRevLett.109.181301
  [arXiv:1207.5988 [astro-ph.CO]].

\bibitem{Queiroz:2015utg}
   F.~S.~Queiroz and C.~E.~Yaguna,
   JCAP {\bf 1602} (2016) no.02,  038
   doi:10.1088/1475-7516/2016/02/038
   [arXiv:1511.05967 [hep-ph]].

  \bibitem{4jmet}
  A. Ilnicka, S. Moretti, P. Poulose, T. Robens, Shibananada Sahoo, {\em In Preparation}.

 \bibitem{Alwall:2014hca}
   J.~Alwall {\it et al.},
   JHEP {\bf 1407}, 079 (2014)
   doi:10.1007/JHEP07(2014)079
   [arXiv:1405.0301 [hep-ph]].

  \bibitem{feynrules}
   http://feynrules.irmp.ucl.ac.be/wiki/InertDoublet
   
\bibitem{Sjostrand:2006za} 
  T.~Sjostrand, S.~Mrenna and P.~Z.~Skands,
  JHEP {\bf 0605}, 026 (2006)
  doi:10.1088/1126-6708/2006/05/026
  [hep-ph/0603175].
  
\bibitem{Conte:2012fm} 
  E.~Conte, B.~Fuks and G.~Serret,
  Comput.\ Phys.\ Commun.\  {\bf 184}, 222 (2013)
  doi:10.1016/j.cpc.2012.09.009
  [arXiv:1206.1599 [hep-ph]].





  

  
\end{thebibliography}
\end{document}